\documentclass{iaus}
\usepackage{graphicx}
\usepackage{natbib}

\newcommand{\apj}{ApJ}
\newcommand{\apjs}{ApJS}
\newcommand{\apjl}{ApJL}
\newcommand{\aap}{A{\&}A}
\newcommand{\aaps}{A{\&}AS}
\newcommand{\mnras}{MNRAS}
\newcommand{\aj}{AJ}

\title[Parsec-scale X-ray Flows] 
{Parsec-scale X-ray Flows in High-mass Star-forming Regions}

\author[Townsley \etal ]   
{L. K. Townsley, P. S. Broos, E. D. Feigelson, and G. P. Garmire}

\affiliation{Department of Astronomy \& Astrophysics, 525 Davey
Laboratory, Pennsylvania State University, University Park, PA 16802, USA \break email: townsley, patb, edf, garmire@astro.psu.edu}

\pubyear{2005}
\volume{227}  
\pagerange{?--?}
\date{?? and in revised form ??}
\jname{Massive Star Birth:  A Crossroads of Astrophysics}
\editors{R. Cesaroni, E. Churchwell, M. Felli, \& C.M. Walmsley, eds.}
\begin{document}

\maketitle

\begin{abstract}
The {\em Chandra X-ray Observatory} is providing remarkable new views
of massive star-forming regions, revealing all stages in the life cycle
of high-mass stars and their effects on their surroundings.  We present
a {\em Chandra} tour of several high-mass star-forming regions,
highlighting physical processes that characterize the life of a cluster
of high-mass stars, from deeply-embedded cores too young to have
established an HII region to superbubbles so large that they shape our
views of galaxies.  Along the way we see that X-ray observations reveal
hundreds of stellar sources powering great HII region complexes,
suffused by both hard and soft diffuse X-ray structures caused by fast
O-star winds thermalized in wind-wind collisions or by termination
shocks against the surrounding media.  Finally, we examine the effects
of the deaths of high-mass stars that remained close to their
birthplaces, exploding as supernovae within the superbubbles that these
clusters created.  We present new X-ray results on W51 IRS2E and 30
Doradus and we introduce new data on Trumpler 14 in
Carina and the W3 HII region complexes W3 Main and W3(OH).

\end{abstract}

\firstsection
\section{Introduction}

Most stars are born in massive star-forming regions (MSFRs); the most
massive stars live out their short lives in this environment and
eventually transform it when they explode as supernovae.  In the
meantime, they have a profound influence on their natal neighborhood,
generating HII regions and wind-blown bubbles and often triggering new
generations of stars to form in the surrounding molecular clouds.  The
kinetic power of a massive O-star's winds injected into its stellar
neighborhood over its lifetime equals that input in its supernova
explosion; essentially from the moment they are born high-mass stars
cause changes in their environment on parsec scales.

X-ray observations probe different energetic components of MSFRs than
traditional optical and IR studies.  Stars of virtually all masses and
stages emit X-rays in their youth, although the mechanisms for X-ray
emission vary with stellar mass.  For OB stars excavating an HII region
within their nascent molecular cloud, diffuse X-rays may be generated
as fast winds shock the surrounding media \citep{Weaver77}; we have
recently discovered such diffuse emission with X-ray observations of
M~17 and the Rosette Nebula \citep{Townsley03}.  X-ray studies also
detect the presence of past supernovae through the shocks in their
extended remnants.  

{\em Chandra} and its Advanced CCD Imaging Spectrometer (ACIS) camera
give us the sensitivity, spatial resolution, and broad bandpass to
detect diffuse X-ray emission generated by these high-mass stars and to
separate it from the hundreds of pre-main sequence X-ray-emitting stars
seen in these fields.  {\em Chandra} routinely penetrates heavy
obscuration ($A_V > 100$~mag) with little source confusion or
contamination from unrelated objects to reveal the young stellar
populations in MSFRs.  Before the {\em Chandra} era, the relative X-ray
contributions of high-mass and low-mass stars, OB winds, and supernova
remnant shocks in these regions were largely unknown.

Through the {\em Chandra} General Observer and Guaranteed Time
programs, we are pursuing a multi-year study of MSFRs, cataloguing and
characterizing the point source populations \citep[e.g.][]{Getman05} as
well as searching for diffuse emission.  Last year we reviewed our {\em
Chandra} observations of M17, RCW49, and W51A \citep{Townsley04}.  We
show other examples of our program in this contribution, presenting new
results for W3, W51A IRS2E, Trumpler 14 in Carina, and 30 Doradus in
the LMC.

\section{W3}

W3 is an obscured complex of high-mass stars, H~II regions, and
associated molecular clouds situated 2.3~kpc from the Sun, part of a
vast star formation complex also containing the W4 superbubble, the
massive stellar clusters IC~1805, IC~1795, and NGC~896, and several
unnamed IR clusters \citep{Carpenter00}.  It is bordered to the west by
HB3, a very large, evolved supernova remnant (SNR), which is clearly
seen in the {\em ROSAT} All-Sky Survey image and was observed with {\em
Einstein} \citep{Leahy85}.  Radio studies \citep{Routledge91} suggest
that the SNR shock has not yet reached the W3 H~II regions, but it is
influencing the distribution of CO in the W3 molecular cloud.
\citet{Lada78} and \citet{Thronson85} argued that star formation in W3
is being induced by the expansion of W4, which is sweeping up molecular
gas into a high-density layer, within which stars are forming.
\citet{Oey05} propose that the young (3--5 Myr) OB cluster IC~1795,
triggered to form by W4, is blowing its own second-generation
superbubble at the molecular cloud interface, triggering in turn the W3
MSFR.

\begin{figure}[hbt]
\centering
\includegraphics[width=0.35\textwidth]{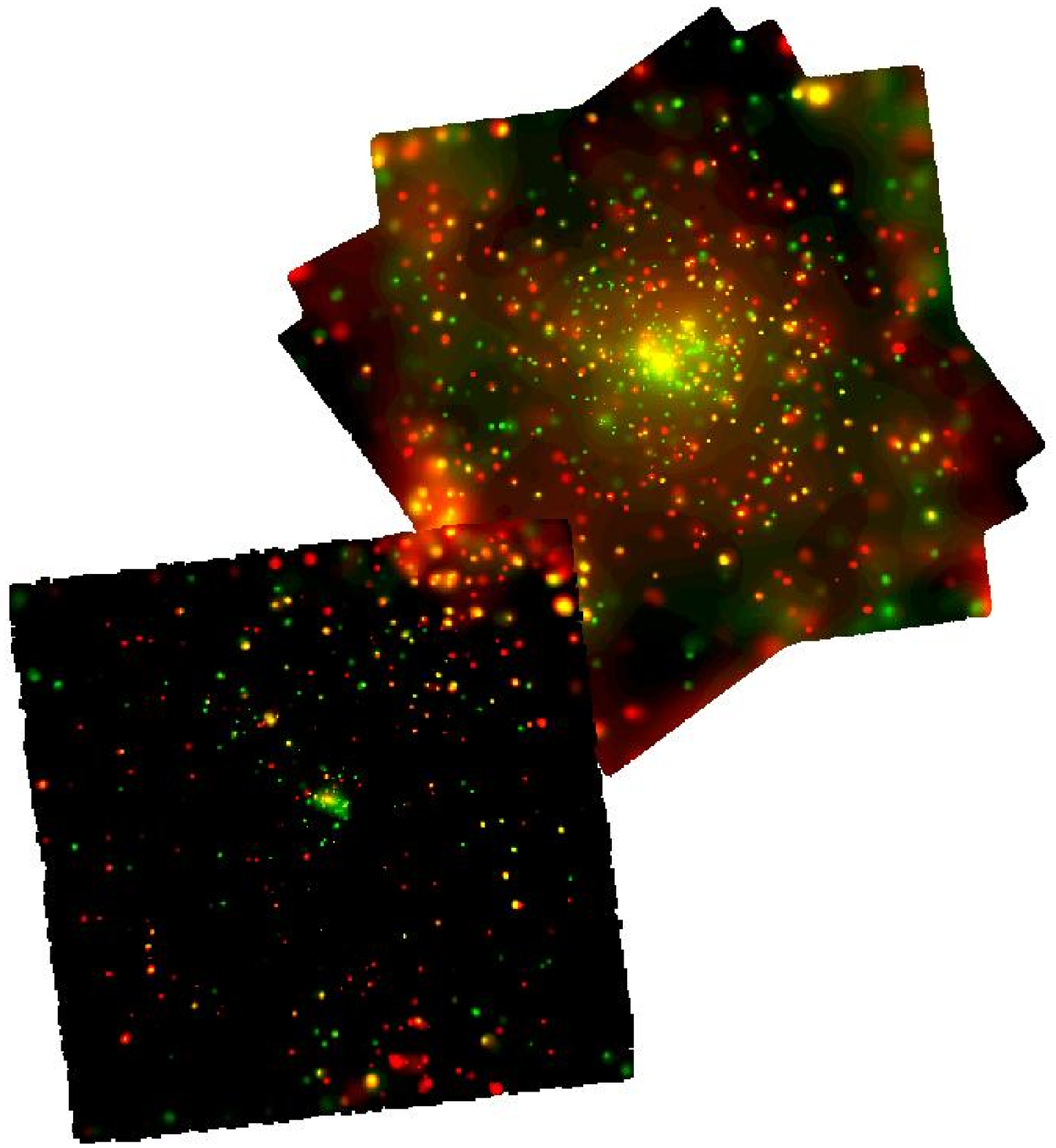}
\protect\hspace{0.5in}
\includegraphics[width=0.35\textwidth]{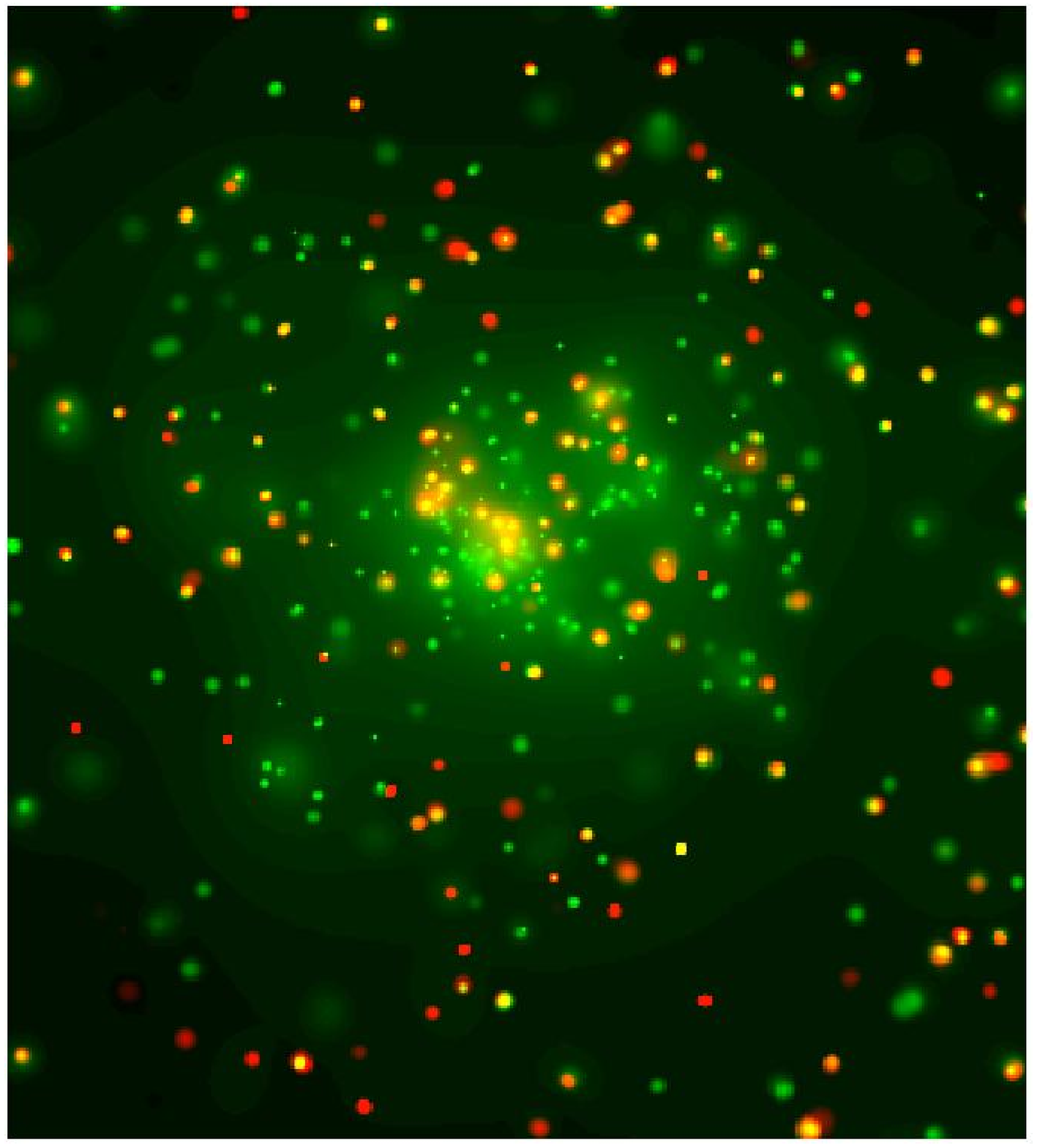}
\caption{
{\bf Left:}  An ACIS-I mosaic of W3 (red = 0.5--2 keV, green =
2--8~keV), including three pointings on W3 Main totaling 78~ks and a
single 72-ks pointing on W3(OH).  Each ACIS-I image is $17^\prime
\times 17^\prime$, or $\sim 11$~pc on a side.
{\bf Right:}  A zoomed image of W3 Main, showing the large number of young, embedded stars revealed by this observation.   
 }
 \label{fig:w3}
\end{figure}

A 40-ksec {\em Chandra} observation of W3 Main using the ACIS imaging
array (ACIS-I) revealed the ionizing sources for many of its HII
regions and over 200 point sources \citep{Hofner02}.  We have recently
obtained more {\em Chandra}/ACIS-I observations of W3 Main and the
adjacent field W3(OH) (Figure~\ref{fig:w3}), showing a rich stellar
population around W3 Main, the older cluster IC~1795 (perhaps
exhibiting soft diffuse emission), and several small embedded clusters
in and around W3(OH).  The W3(OH) field is noticeably lacking in
sources compared to the W3 Main field.  Although partly an obscuration
effect, this also illustrates the intrinsic difference in the size of
these clusters; hard X-rays are largely unaffected by the obscuring
material so the 2--8~keV (green) component in Figure~\ref{fig:w3} reflects
an intrinsic difference in cluster size between W3(OH) and W3 Main.

\section{W51A and the Enigmatic Source IRS2E}

W51 is one of the most massive star-forming complexes in the Galaxy but
is difficult to observe because of its distance ($\sim 7$~kpc) and high
obscuration.  Our 72-ksec {\em Chandra}/ACIS observation of W51A
\citep{Townsley04} detected many of the known radio HII regions
\citep{Mehringer94} as diffuse X-ray sources.  We also see $\sim 450$
point sources, revealing the highest-mass and youngest inhabitants
fueling the HII regions and just emerging from their dusty natal
cocoons.  

\begin{figure}[hbt]
\centering
\includegraphics[width=0.35\textwidth]{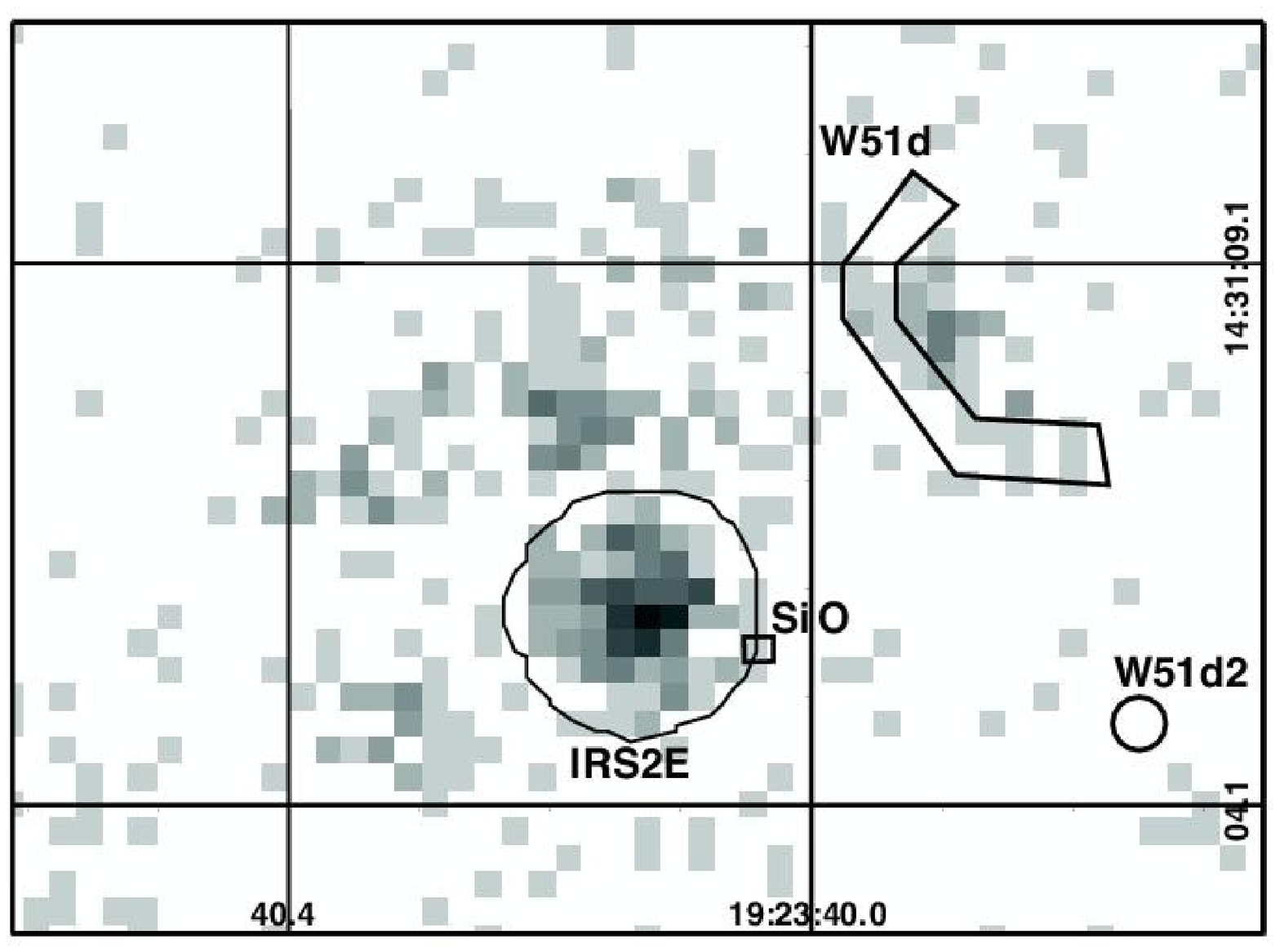}
\protect\hspace{0.5in}
\includegraphics[width=0.35\textwidth]{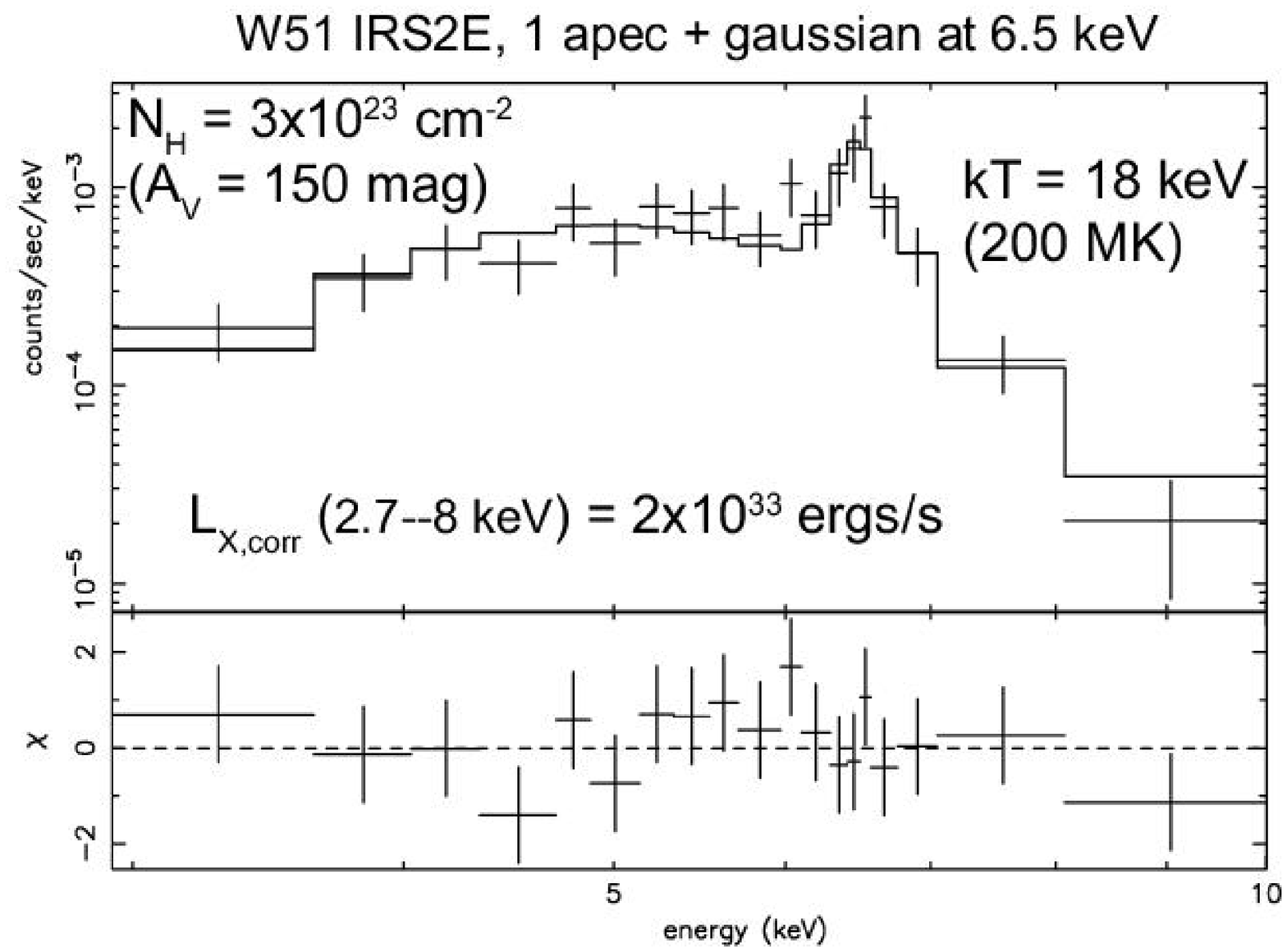}
\caption{
{\bf Left:}  A binned ACIS image of the W51 IRS2 complex,
$8^{\prime\prime} \times 12^{\prime\prime}$, with a J2000 coordinate
grid.  The extraction region for IRS2E, containing 90\% of the 1.5~keV
point spread function, is shown, as is the cometary HII region W51d.
The locations of the UCHII region W51d2 and the strong SiO maser are
noted.
{\bf Right:}  ACIS spectrum of W51 IRS2E:  the upper panel shows the source
spectrum and model fit; the lower panel shows the fit residuals.
} 
\label{fig:w51}
\end{figure}

Buried in one of the youngest and richest high-mass complexes,
G49.5-0.4, we have discovered an enigmatic hard X-ray source at the
center of an embedded high-mass stellar cluster
(Figure~\ref{fig:w51}).  CXOW51~J192340.1+143105 is spatially
coincident with a deeply-embedded mid-IR source \citep{Kraemer01} known
as IRS2 East (IRS2E).  It is surrounded by powerful masers and
ultra-compact HII regions yet has no associated radio HII region
itself.  Evidence for infall is seen in this region \citep{Sollins04}. 

IRS2E emits most of its X-ray photons in a broad 6.5-keV line probably
due to fluorescent iron.  Although common in AGN, this type of X-ray
spectrum is highly unusual for stellar sources, as it requires an
embedded source with substantial emission above the iron absorption
edge (7.1~keV).  Such a hard X-ray spectrum could be generated by
colliding winds in a massive cluster \citep{Canto00}, but the X-ray
emission of IRS2E varies by a factor of two in 40~ksec, ruling out the
cluster explanation.  We suspect that it is a colliding wind binary,
perhaps a younger version of Eta Carinae \citep{Corcoran04} or HD~5980, a
luminous blue variable in the SMC \citep{Naze02}.  The absence of an
HII region around this source suggests that it is very young.

\section{Trumpler 14 in Carina}

The Carina complex, at a distance of $\sim$2.8~kpc \citep{Tapia03}, is
a remarkably rich star-forming region at the edge of a giant molecular
cloud (GMC), containing 8 open clusters with at least 64 O stars, 2
Wolf-Rayet stars, and the luminous blue variable Eta Carinae
\citep{Feinstein95}.  {\em ISO} discovery of 22$\mu$m grains in the
bright radio HII region Carina I may imply that a supernova occurred in
this region \citep{Chan00}; the presence of WR stars also may indicate
past supernovae, although no well-defined remnant has ever been seen.

Tr~14 is an extremely rich, young ($\sim$1~My), compact OB cluster near
the center of the Carina complex, containing at least 30 O and early B
stars \citep{Vazquez96}.  Tr~14 is probably at the same distance as its
neighboring, equally rich cluster Trumpler~16 but is thought to
be younger \citep{Walborn95}.  These two clusters contain
the highest concentration of O3 stars known in the Galaxy; their
ionizing flux and winds may be fueling a bipolar superbubble
\citep{Smith00}.

An {\em Einstein} X-ray study of the Carina star-forming complex was
performed by \citet{Seward82}.  They detected $\sim$30 point sources,
mostly individual high-mass stars and the collective emission from
unresolved cluster cores.  They also detected diffuse emission
pervading the entire region and speculated that it may be due to O star
winds.  Based on experience with {\em Chandra}, we now know that
thousands of the lower-mass stars in these young clusters were likely
to be contributing to the diffuse flux seen in the {\em Einstein}
data.  A major goal of our {\em Chandra} observation was to resolve out a
significant fraction of this point source emission so a better
determination of the spatial and spectral characteristics of the
diffuse component can be made.

\begin{figure}[hbt]
\centering
\includegraphics[width=0.32\textwidth]{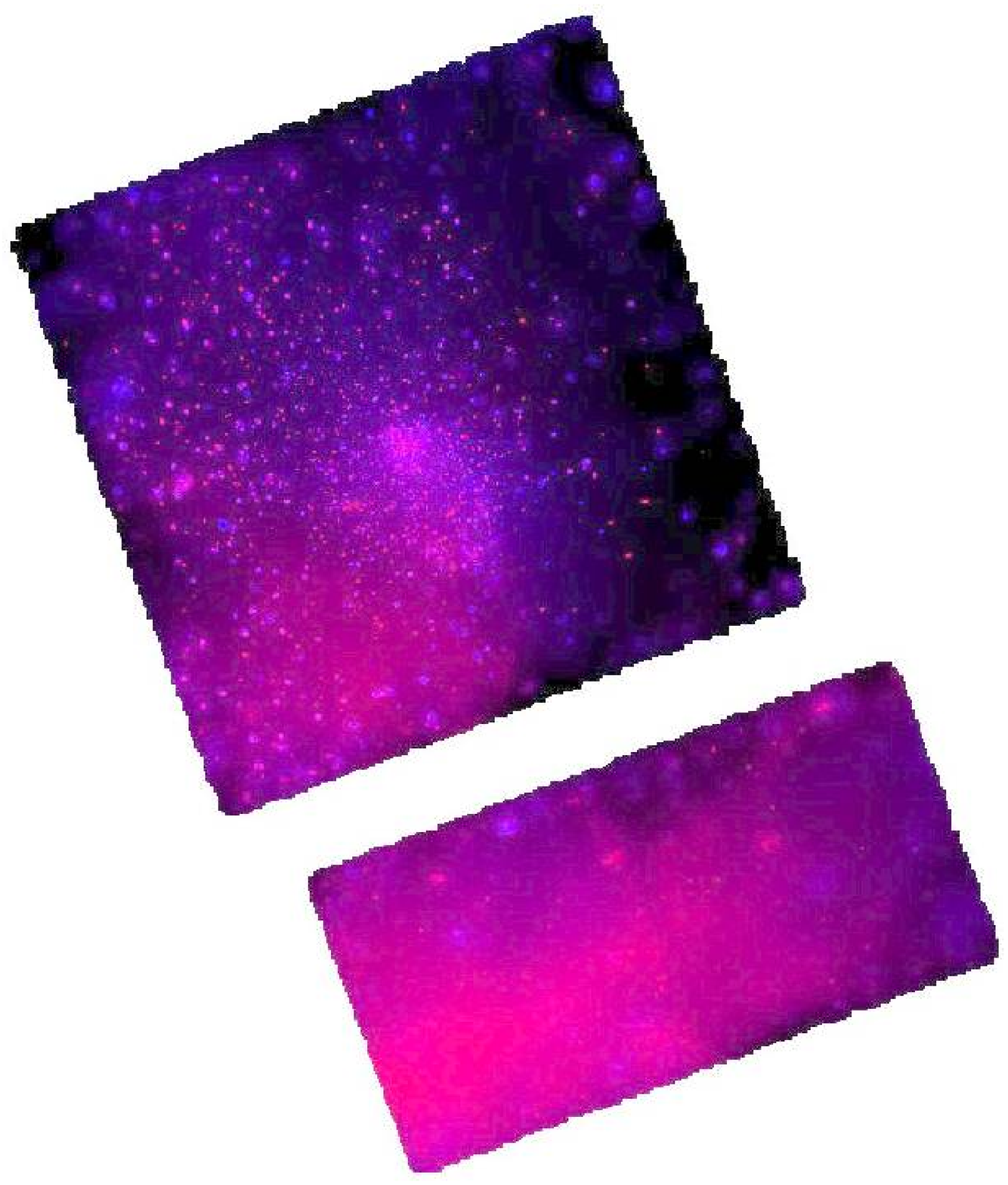}
\protect\hspace{0.8in}
\includegraphics[width=0.45\textwidth]{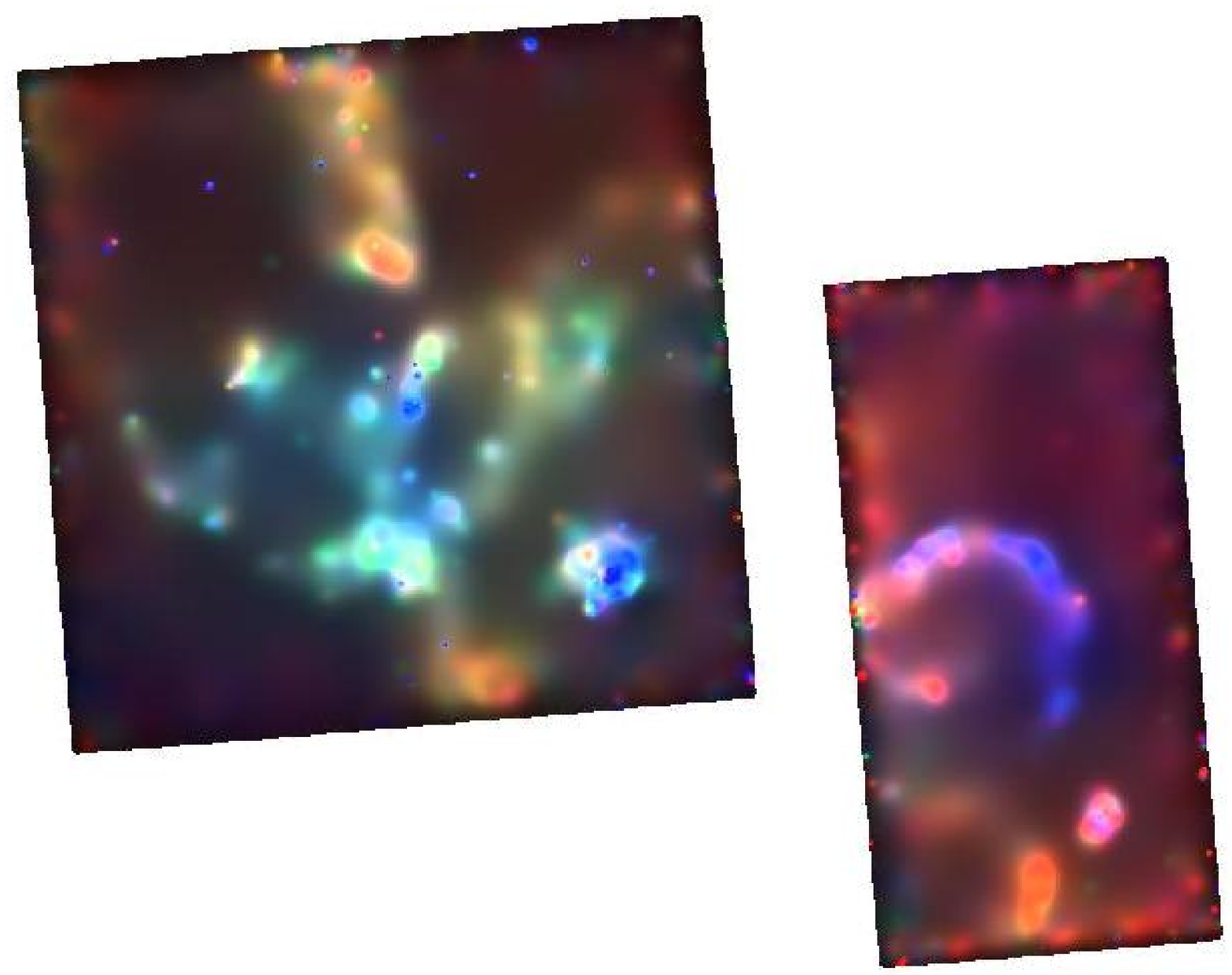}
\caption{
{\bf Left:}  A smoothed image (red = 0.5--2~keV, blue = 2--8~keV) of
the 57-ksec ACIS observation of Tr~14 in Carina, with the cluster
imaged on the ACIS-I array ($17^\prime \times 17^\prime$, or $\sim
14$~pc on a side at D = 2.8~kpc) and bright diffuse emission seen in
the off-axis S2 and S3 CCDs (each $8.5^\prime \times 8.5^\prime$).  We
find $\sim 1600$ point sources on the I array plus extensive diffuse
emission across the whole field.
{\bf Right:}  Smoothed soft-band image (red = 0.5--0.7~keV, green =
0.7--1.1~keV, blue = 1.1--2.3~keV) of the 21-ksec ACIS observation of
30~Doradus \citep[from][]{Townsley05a}, with 30~Dor Main imaged on the
I array (covering $\sim 250$~pc on a side at D = 50~kpc); the large
shell 30~Dor~C, the Honeycomb SNR, and SN1987A are seen in the off-axis
CCDs S3 and S4.  Bright, soft diffuse emission dominates the entire
field.
} 
\label{fig:cavity}
\end{figure}

The aimpoint of our 57-ksec ACIS observation of Tr~14
(Figure~\ref{fig:cavity} left) was the central star in the cluster,
HD~93129AB, a very early-type (O2I--O3.5V) binary \citep{Walborn02},
with the two components separated by $\sim 1^{\prime\prime}$.  Since
these are resolved in the ACIS data, we can see that the two components
have very different spectra; HD~93129B shows a typical O-star X-ray
spectrum ($kT = 0.5$~keV, or $T \sim 6$~MK), while HD~93129A shows a
similar soft component ($kT = 0.6$~keV) but also exhibits a much harder
component, with $kT = 3.0$~keV ($T \sim 35$~MK), and is ten times
brighter in X-rays than HD~93129B.  This hard spectrum and high X-ray
luminosity are indicative of a colliding-wind binary \citep{PPL02}; in
fact HD~93129A was recently discovered to be a spectroscopic binary
\citep{Nelan04}.  Additionally, while the O3V star HD~93128 is soft and
fainter in X-rays, we find that the O3V star HD~93250 in Tr~14 shows a
two-component spectrum and X-ray luminosity almost identical to
HD~93129A.  {\em XMM-Newton} observations also show a hard spectral
component for HD~93250 \citep{Albacete03}; this source is very likely a
colliding-wind binary as well.

The diffuse emission in Tr~14 is quite soft and shows abundances
typical of OB wind termination shocks.  We also see soft, bright
diffuse emission in the off-axis CCDs of the ACIS array, far from any
of the Carina massive stellar clusters.  Spectral fits to this diffuse
emission require abundances of O, Ne, Si, and Fe to be more than twice
the solar value; this is evidence that the emission may be from an old
``cavity'' supernova remnant that exploded inside the Carina
superbubble, as suggested by \citet{Chu93}.

\section{30 Doradus}

Early in the {\em Chandra} mission, we obtained a $\sim 21$~ksec
observation of the most luminous Giant Extragalactic HII Region and
``starburst cluster'' in the Local Group, 30~Doradus in the Large
Magellanic Cloud.  The ACIS pointing was centered on the young, dense
OB cluster R136, a testbed for understanding recent and ongoing star
formation in the 30~Dor complex.  The presence of evolved supergiants
$\sim 25$~Myr old and embedded massive protostars shows that 30~Dor is
the product of multiple epochs of star formation, including a new
generation of embedded stars currently forming, possibly as a result of
triggered collapse from the effects of R136 \citep{Brandner01}.
Supernovae pervade the region but may go undetected due to age and
environment \citep{Chu90}.  Nearby are two msec pulsars and SN1987A.
30~Dor produced at least five plasma-filled superbubbles with
$\sim$100-pc scales \citep{Wang91}, likely products of strong OB winds
and multiple supernovae.

The right panel of Figure~\ref{fig:cavity} shows a smoothed ACIS image
from our 21-ksec observation of 30~Dor, including the off-axis CCDs S3
and S4 as well as the main 30~Dor nebula on the ACIS-I array.  We see a
bright concentration of X-rays associated with the R136 star cluster,
the bright SNR N157B to the southwest, a number of new
widely-distributed compact X-ray sources, and diffuse structures
associated with the superbubbles produced by the collective effects of
massive stellar winds and their past supernova events
\citep{Townsley05a}.  Some of these are center-filled while others are
edge-brightened, indicating a complicated mix of viewing angles and
perhaps filling factors.

Our spectral analysis of the superbubbles reveals a range of
absorptions ($N_H = 1$--6$ \times 10^{21}$~cm$^{-2}$), plasma
temperatures ($T = 3$--9$ \times 10^6$~K), and abundance variations.
We find $\sim 100$ sources associated with the central massive cluster
R136 \citep{Townsley05b}; some bright, hard X-ray point sources in the
field are likely colliding-wind binaries \citep{PPL02}.  Comparing the
X-ray data to visual and IR images, we find that hot plasma fills the
shells outlined by ionized gas and warm dust.

\section{Summary}

Interactions between powerful O star winds and the ISM lead to
parsec-scale soft X-ray emission as predicted by \citet{Weaver77} and
others, but with much fainter X-ray luminosities; this hot plasma may
pervade the Galactic plane but is hard to detect due to obscuration.
Wind-wind interactions lead to harder X-rays; this emission may provide
a way to determine close binarity in massive stars or to detect
embedded massive clusters.  The $10^4$K Str{\" o}mgren Sphere that
defines classical HII regions is really a Str{\"o}mgren Shell filled
with $10^6$K plasma in many MSFRs.  Only a small portion of the wind
energy and mass appears in the observed diffuse X-ray plasma, though;
it could be dissipated via turbulence, mass-loading, or fissures into
the ISM \citep{Townsley03}.  We see bright, soft diffuse X-rays in some
regions; enhanced metallicity and luminosity compared to wind-generated
emission and the presence of these structures in regions that also
contain evolved stars implies that these X-ray features are the remains
of cavity SNRs.

{\em Chandra} chronicles the life cycle of massive stars through
studies of MSFRs.  In W3 and W51A we see massive embedded protostars;
winds from main sequence O stars fill M17 and Tr14 with soft X-rays.
Colliding-wind binaries appear as hard X-ray sources in Tr14, W51
IRS2E, and 30~Dor.  Cavity supernova remnants dominate the soft diffuse
X-ray emission in Carina and 30~Dor, enhancing the superbubbles blown
by the winds of massive stars.

\begin{acknowledgments}
This work was supported by the National Aeronautics and
Space Administration (NASA) through contract NAS8-38252 and
{\em Chandra} Awards G04-5006X, G05-6143X, and SV4-74018 issued by the
{\em Chandra} X-ray Observatory Center, operated by the Smithsonian
Astrophysical Observatory for and on behalf of NASA under contract
NAS8-03060.

\end{acknowledgments}

\end{document}